**The GAIN Model: A Nature-Inspired Neural Network Framework Based on an Adaptation of the Izhikevich Model**

Gage K. R. Hooper

Independent Researcher

Future Aerospace Engineering Student, Embry-Riddle Aeronautical University

May 31, 2025

**Abstract**

While many neural networks focus on layers to process information, the GAIN model uses a grid-based structure to improve biological plausibility and the dynamics of the model. The grid structure helps neurons to interact with their closest neighbors and improve their connections with one another, which is seen in biological neurons. While also being implemented with the Izhikevich model this approach allows for a computationally efficient and biologically accurate simulation that can aid in the development of neural networks, large scale simulations, and the development in the neuroscience field. This adaptation of the Izhikevich model can improve the dynamics and accuracy of the model, allowing for its uses to be specialized but efficient.

**Literature Review**

**Historical Development of Spiking Neural Networks**

Spiking Neural Networks (SNNs) are a biologically inspired neural network[1] that can model neuron behaviors closer than traditional artificial neural networks. SNNs communicate through the timing and frequency of spikes[2] then information is encoded through it. Early models of SSNs, such as the Hodgkin-Huxley model (1952), were detailed and capable of replicating the exact dynamics of neuronal spiking, considering every ion channel, but it was too computationally inefficient. More computationally efficient alternatives, such as the Leaky

---

[1] A computational model that can simulate the function of neurons.
[2] The activation of neurons determined by its action potential when a neuron's difference between interior and exterior voltages (membrane potential) rapidly increases and decreases.

Integrate-and-Fire (LIF) model, simplified the dynamics, but lacked the ability to replicate more complex and dynamic behaviors like bursting[3] or adaptation.

In response to limitations seen in these models, Eugene Izhikevich (2003) introduced a spiking neural network model, achieving a balance between biological plausibility and computational efficiency (See Appendix A). The Izhikevich model can reproduce neuron behaviors while remaining computationally lightweight, resulting in it being widely adopted for large-scale simulations. The model's simplicity is one of its largest advantages, but this simplification causes certain neural functions to be completely forgotten.

## Limitations of Existing Models in Capturing Neuronal Adaptation

One of the limitations of the Izhikevich model is that the model's simplification is unable to accurately represent the adaptability that is present in real-world neuronal behavior, such as current-based adaptation and spike-frequency adaptation (SFA)[4]. The static input current helps the stability of the neural network. Biological neurons, however, adapt their firing rate and membrane potentials[5] by responding to varying input currents. This is especially important where the neurons are exposed to fluctuating inputs over time. Benda and Herz (2003) showed that SFA plays a crucial role in sensory processing and neuronal stability. However, SFA tends to be underrepresented in the simple spiking neuron model, including the Izhikevich model.

---

[3] A general behavior that describes periods that neurons fire rapidly which are then followed by much longer than typical inter-spike intervals.

[4] The behavior described by a neuron decreased activity from a sustained current, allowing it to adapt to the pre-synaptic currents. This behavior's complete definition and framework is complicated, but can be defined from Benda and Herz (2003)

[5] The difference in electric potentials in the interior and exterior walls of a cell's membrane

Inspired by the foundations of the Izhikevich model, Grid-based Adaptable Izhikevich Network (GAIN), my model addresses his model's limitations. Modifying it to handle SFA for more complex neuronal interactions, while being able to handle dynamic and changing inputs.

**Current-based Adaptation in Neuronal Models**

Current-based mechanisms are essential for accurately modeling the dynamic behaviors of biological neurons. It allows the neurons to adjust their membrane potential, and firing patterns based on sustained inputs. It is critical for understanding biological processes like adaptation and habituation[6]. Research by Marder et al. (2002) and Koch (1999) has shown that when neurons adapt in response to input current, it can significantly influence the spiking patterns of neurons, specifically in sensory systems.

Despite these findings, most computational models ignore or oversimplify it, resulting in inaccurate simulations of neurons within a changing environment. The GAIN model introduces current-based adaptation, allowing it to dynamically adjust its membrane potential based on the input current over time, which improves biological realism and the network stability. This modification improves the Izhikevich model's handling of long-term input currents without increasing computational complexity.

**Synaptic Plasticity and Spike-Timing Dependent Plasticity**

Synaptic plasticity, specifically spike-timing dependent plasticity (STDP, See Appendix C), is a crucial learning mechanism in neural networks. Where the timing of pre- and post-synaptic spikes[7] determines if the synapses strengthen (long-term potentiation, LTP) or weaken (long-term depression, LTD), as described by the study of Bi and Poo (2001). The Izhikevich model accounts for mechanisms for plasticity controlled by current adaptation but does not specifically include it on its own. Which can play a role in more complex behaviors, like memory formation tasks and pattern recognition, as seen in biological neurons.

_______________

[6] Neurons decreased behavioral responding when repeatedly exposed to a stimulus.

Recent studies, such as Zenke et al. (2015), have shown that combining both current-based dynamics with STDP can lead to improved learning behaviors in neural networks. By integrating current-based adaptation, the GAIN model improves the accuracy of synaptic plasticity, particularly in response to long-term stimuli, without compromising computational efficiency.

**Grid-Based Neural Networks**

The GAIN model improves on the original framework of SNN models, where it is a current-based adaptation paired with grid-based structure. Each neuron has its own separate cell within the grid. The advantage of using STDP integrated into the model is where each neuron is connected to its close neighbors (See Appendix B1). To increase the number of connections per neuron, you increase the number of dimensions[8]. A three-dimensional grid has twenty-six neighboring neurons per neuron, and it continues to higher dimensions. The benefit is having a completely connected network, and the limited number of neighbors encourage the connection between each cell. Encouraging memory and pattern recognition, simplifying the structures seen in nature, while making it computationally lightweight.

---

[7] When a neuron sends a signal it is considered the pre-synaptic neuron, while the one that receives the signal is the post-synaptic neuron.

[8] The dimensions of the matrix, which represents the grid, more layers can be added to it to increase complexity.

The technique to interpret and input information to the neurons is closely related to vitro neural networks, cultures of neurons integrated into a multielectrode array (MEA)[9], where the outputs of neurons can be interpreted by the response of the neurons to the stimuli. For this study, to replicate MEA's instead of the physical structure of the brain, the model would be two dimensional. From this, the GAIN model is a dynamic network that is biologically plausible, computationally efficient, adaptable, which can have uses impossible for traditional neural networks, and extremely difficult for simple SSNs.

**Applications for GAIN**

GAIN's adaptability to stimuli and being able to respond to it dynamically offers a variety of uses. As memory and pattern recognition can be used within robotics, its responses from stimuli with a current-based system can be paired with audio devices for sound recognition, or even reconstruct audio from listening to sounds. This also applies to other sensory information, such as visual information. With its benefit of being computationally efficient, it can also be used for large scale biological simulations. The use of these biological simulations can be used to study systems within the brain and decrease animal experimentation in neurology. Its applications can be anything that has uses needed for pattern recognition, and a balance between biological accuracy and computational efficiency.

**Hypothesis**

From what has been previously mentioned the reason for this study is to answer if the Izhikevich model can be adjusted to be more biologically plausible and to implement a grid-based structure to improve the interactions between neurons. A fundamental inspiration for the gird-based structure was neuron simulations that had physical space between neurons and simulated the structure of the brain and its connections. The grid-based structure simplifies it so that all the neurons' data is side by side, and the distance between each neuron can still be simulated by weight. This is a large improvement in traditional neural networks that have a more linear structure, where the neurons are connected by layers that move the signal forwards. The grid-based structure allows every neuron to be connected to one another and be able to interact.

---

[9] An array of multielectrode that can send and receive neural signals acting as an interface between computers and neurons.

## Methodology

### Structure of the Model

To adapt the Izhikevich model to be more dynamic and accurate, I altered the model to take the input current into account and affect the recovery variable of the neuron. I also increased the dynamics of the model by implementing STDP and short-term synaptic plasticity (STP, See Appendix D), allowing the model to act more biologically accurately. To simulate the connections between neurons and their neighbors mimicking the biological structure of neurons, I made each neuron an element within a matrix where each neuron only interacts with its closest neighbors. This encourages the use of synaptic plasticity within the model, and theoretically improving the pattern recognition and memory of the system. Inspired by vitro neurons, I implemented the matrix structure of the model into a grid, where the inputs and outputs will be like MEAs. I also adjusted the variables that are part of the recovery variable to be their own functions of input current and membrane potential. With another two constants considered for each that can adjust how reactive the neurons are to a stimulus. (See Appendix B)

**Implementing and Interpreting GAIN**

Like vitro neural networks, interpreting the outputs of the model can be related to how the model's activity changes with the input current. I will set specific neurons to be the input and output (I/O) neurons. This can be a specific side, or a pattern set within the grid. Allowing the model to be more useful enabling it to adapt to specific data and simulations, such as a sensory system simulation and use it to recognize specific objects or patterns within media like video or music.

I will first train the network to behave closer to biological neurons, specifically prefrontal cortex (PFC) neurons from data and behaviors that are seen from it. Finding an average range for each variable that the neuron is dependent on can serve as a foundation for what is stable and accurate. This allows the model to be predictable and accurate, which will help with potential future issues, such as the membrane potential increasing to infinity or oscillating unrealistically.

Interpreting the model's biological plausibility will be based on several factors. First seeing if the model includes the three main parts of a neuron's action potential. Depolarization, where the membrane potential rapidly increases. Repolarization, where the neuron's membrane potential tries to recover back to normal and dips below the resting potential, and the refractory period where the neuron's membrane potential fully adjusts back to the resting potential. Next, I will observe if the neurons' membrane potential stays within a reasonable range of -100mV to 40mV which both are already quite high extremes of biological neurons but can still be seen as reasonable. Seeing how often the neurons spike is another large factor, this is because biological neurons have a refractory period. As previously mentioned, it is a function of repolarization where the neuron's membrane potential dips below the resting potential. This dip is what causes the neuron to not be able to spike until it reaches back to its resting potential. Therefore, if a neuron spikes extremely rapidly in only a few milliseconds is too unrealistic.

**Simulations**

The first simulation will be looking for how the model behaves through comparing the behavior to biological neurons, this will help to gauge the effectiveness of this model being a balance between biological plausibility and computational efficiency. The second simulation will expand on describe the activity of the network and how it relates to biological neural networks. This allows me to show data that is not clear within the membrane potential graphs. The third simulation will compare the single neuron behavior with other SNN's such as the Izhikevich model, LIF, and Hodgkin-Huxley model. The Hodgkin-Huxley model will show the behavior of a biological neuron. LIF will show the fastest option for SNN's. Izhikevich will show the current best option for simulating biological neurons quickly and accurately. All the simulations have been programmed with the programming language Python.

While developing the model I am including a noise function and refractory period[10] not included within the model itself but helps the biological plausibility of the simulation. I have also assigned the initial variables of the neurons randomly within a range that is respective for its function to be biologically realistic (Shown by Appendix Z). Implementing the previously mentioned mechanisms I created a simulation for a single spike (Shown by Figure 1). The spike contains the three main parts of a neuron's action potential. Although needing improvements, the spike is biologically accurate enough for training and testing the model.

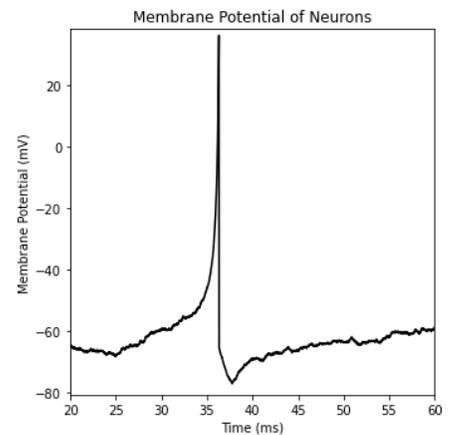

*Figure 1. Singular spike over the intervals 20ms and 60ms*

---

[10] Refractory period referred as the period before a neuron can spike again, this behavior is seen within biological neurons. Without a refractory period, the simulation could spike rapidly and unrealistically.

**Training**

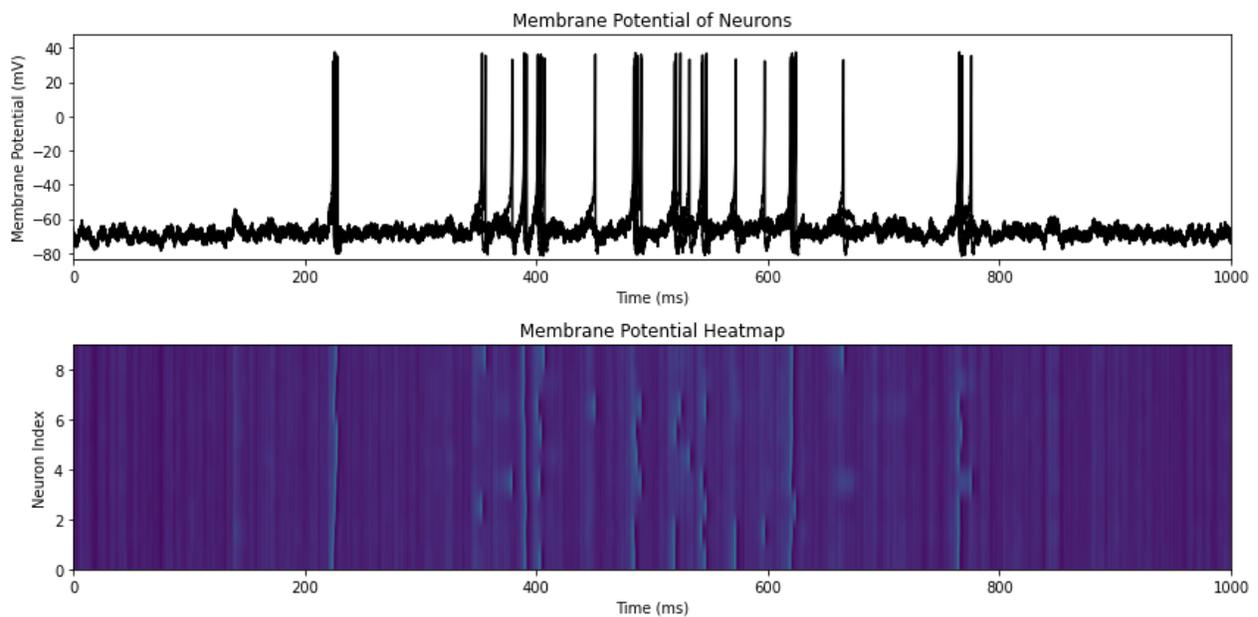

*Figure 2. A simulation of 9 randomly set neurons over a period of 1000ms. The top graph representing the membrane potential of each neuron and the lower graph representing neuron activity as a heatmap.*

The goal of the model being more biologically plausible as compared to the Izhikevich model has been proven by the previous figures. However, the neurons' behaviors tended to be highly active, I described this behavior as chatty neurons, this is associated with low weights between is neighboring neurons. Chatty neurons are more comparable to biological neurons, but I wanted to gauge the effectiveness of training the model with something simple. I trained the model to be less active by using gradient descent (Courant, 1943) to adjust the weight and decrease the number of spikes. This is an incredibly simple test, to ensure that the optimization of the model is working and how the network interacts with one another as its being optimized. I ensured that STP and STDP were working with the model checking the interactions with neighbors and the behavior of a neuron with repeated stimulus (See Appendix Ca and Da). I began by making a simulation of 25 random chatty neurons (Shown by Figure 4a).

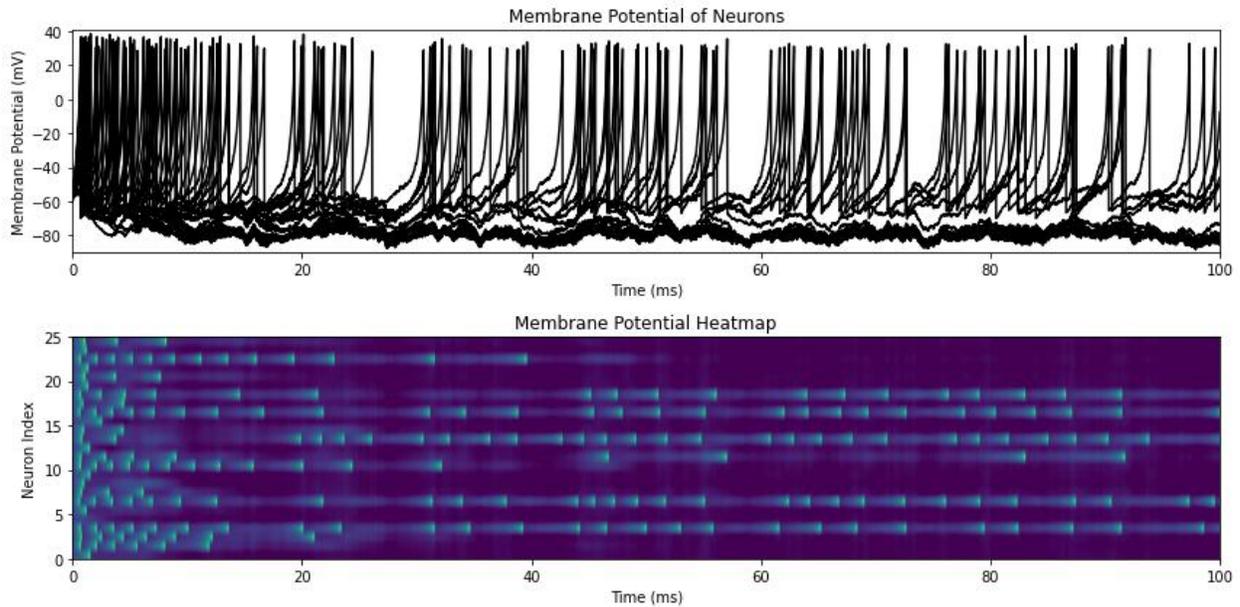

*Figure 4a. A simulation that contains twenty-five random neurons, five of which are chatty.*

I repeated the simulation with the optimization algorithm implemented to decrease the activity. After around 40ms the activity in the network completely stopped (Shown by Figure 4b).

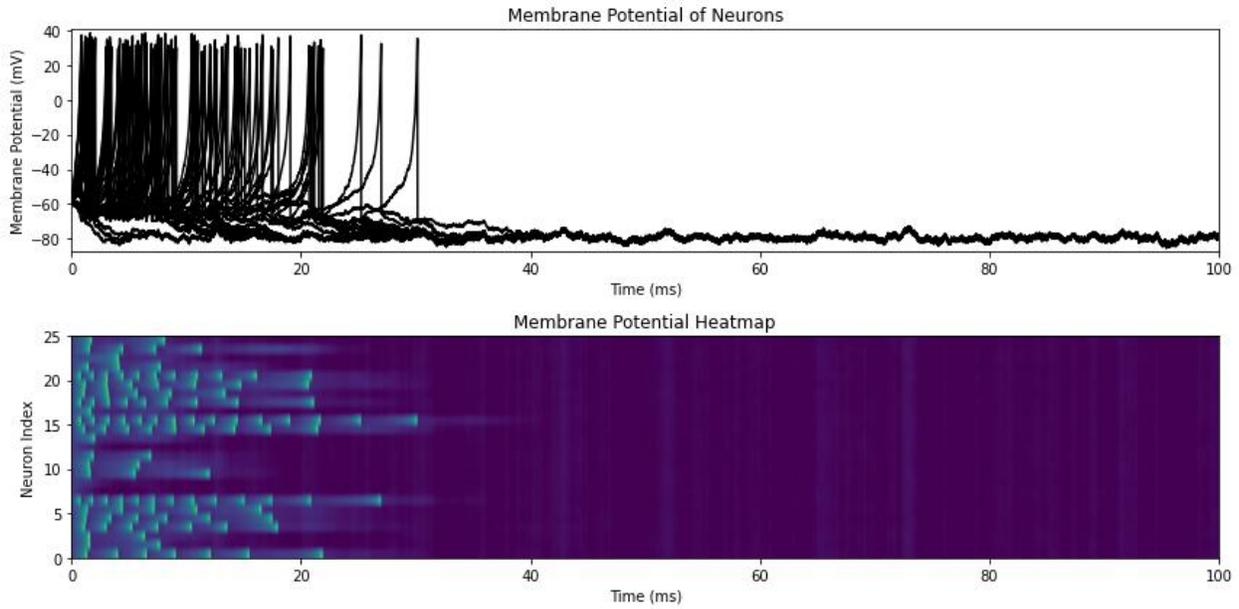

*Figure 4b.*

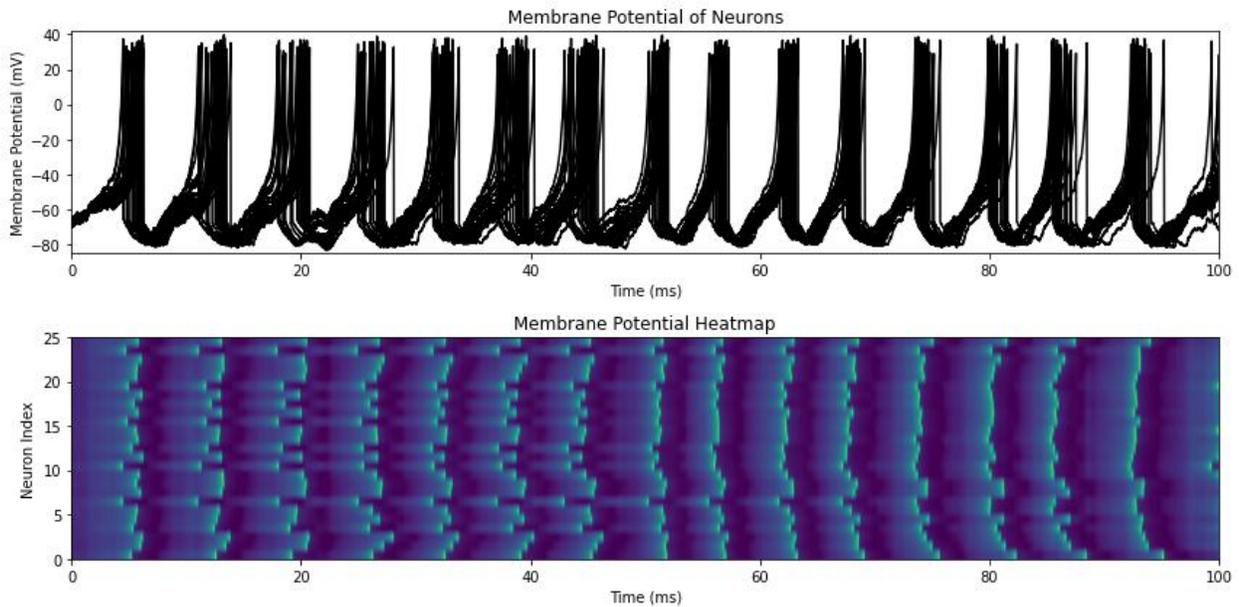

*Figure 5. A set of 25 neurons in a 5 by 5, over the 100ms simulation time with 15mV input across all neurons.*

This small test helps prove the ability for the model to be optimized like traditional neural networks. I also ran more tests to assess the interactions between a neuron and its neighbors, leading to behaviors seen in nature, such as synchronizing neurons (Shown by Figure 5). In biology when neurons synchronize it is a sign that a strong connection between the neurons have been made, shown by the study of Galuske, Munk, and Singer (2019). This helps prove how strong the connections from STDP have formed.

While the network begins slightly chaotic, the neurons over time begin to synchronize with one another, this helps ensure that STP is working. The warping begins to occur, I believe is a result of the grid structure of the model. Where at the beginning all the neurons spike instantaneously, this is from the neurons' first interaction with any kind of stimulus where they need to get adjusted to the current environment. However, overtime the neurons begin to adjust and interact with one another. Since the neurons in the middle have the greatest number of neighbors and the neurons on the corners and edges have the least amount. The neurons in the middle of the system have a larger chance of affecting the rest of the model. This can be seen as the outputs of the neuron propagating throughout the rest of the system. Where with one spike, it causes the neighbors to spike, which then causes their own neighbors to spike, and this process continues until it reaches all the neurons. This behavior of signals propagating throughout the system is extremely reminiscent of biological neurons. This is an extremely good sign that STDP is working and that the connections between the neurons can be seen as biologically accurate.

## Results and Analysis

### First Simulation Setup

The first simulation was conducted with a grid of 484 neurons arranged in a 22x22 format, each neuron initialized with parameters in the range of biological neurons (See Appendix Z). The simulation ran for 250ms with 0.01ms time steps, during which the input current was generated with a constant input of 20mV across the entire grid.

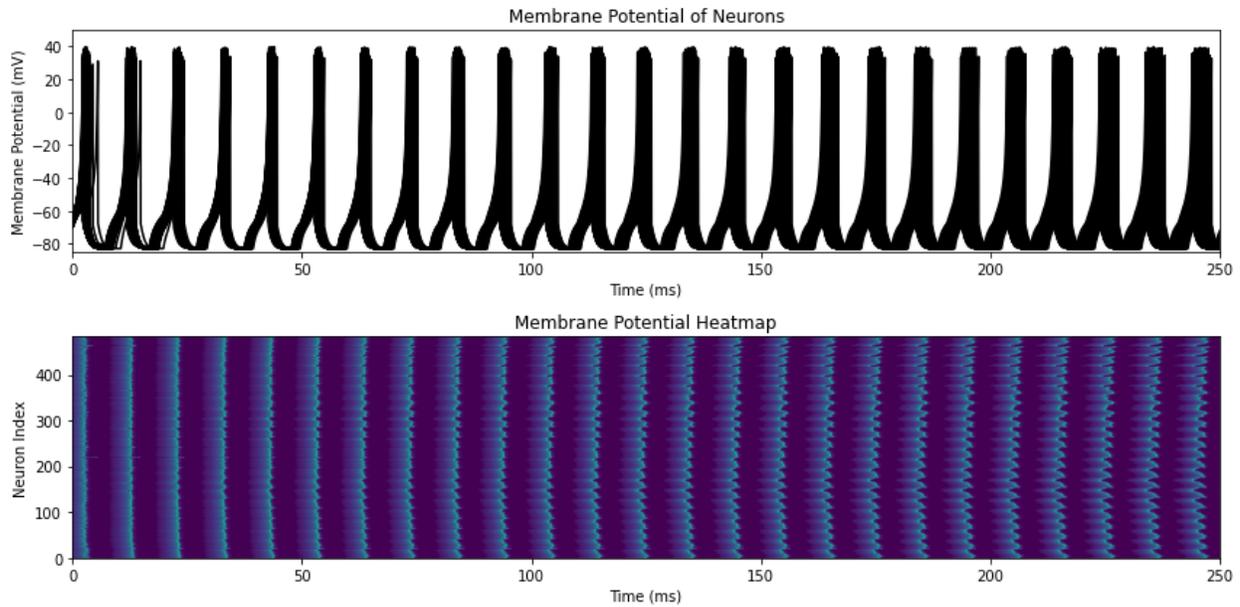

*Figure 6.*

Figure 6 illustrates the membrane potential of the neurons over the simulation time. Expectedly the neurons exhibited spiking behavior from the stimuli. The spikes at the beginning were at first chaotic, but quickly synchronized with one another. Nearing the end the neurons spikes separate this shows the excitability in the network decreasing and the neurons are interacting less. This decrease in activity causes neurons to be more affected by their neighbors, this is what causes the warping to begin and being able to see clear sections between every 22 neurons.

## Synaptic Weight Adaptation

The synaptic weights adjusted throughout the simulation, as shown by Figure 7. Weights were set randomly, but after a series of spikes, weights between neurons that frequently interacted increased, demonstrating synaptic strengthening. Conversely, the weights between neurons that did not interact often diminished, reflecting a basic form of Hebbian learning[11]. This adaptation suggests that the model has implemented the mechanisms of synaptic plasticity, which is crucial for learning and memory in biological neural networks. From this visualization of the grid, it is apparent the strong connections the neurons have to one another and how they affect each other. The reason the change in weights is the best way to visualize the activity of the network is because the neighbors influence on each other is a variable that is changing throughout the entire simulation and describes the connections of the neurons. This is why there are very clear active and inactive spaces in the network even in a small simulation.

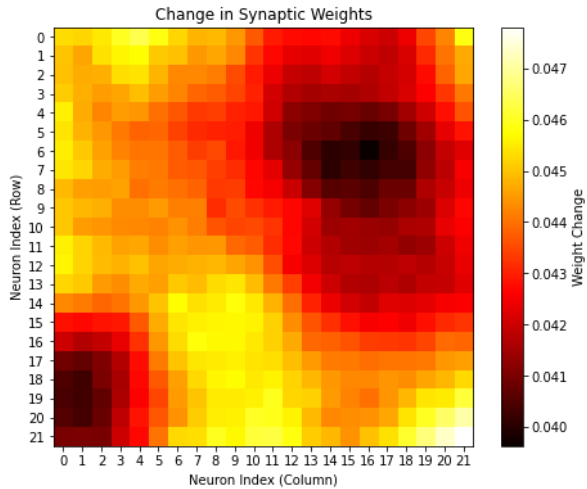

## Results of the Simulation

I found the final synaptic weight distributions. The distribution of synaptic weights followed a power-law distribution (See Figure 8) which is often in biological neural networks, which heavily supports the biological plausibility of the model. From the previous sections we can find the neurons connected extremely well to one another showing visible areas of activity and inactivity. The second simulation goes further into detail of this behavior, where it can be hypothesized that the neurons will separate into their own communities like biological neural networks.

---

[11] First proposed by Donald Hebb (1949), Hebbian learning is a form of activity-dependent synaptic plasticity, where neurons that frequently interact strengthens their connection, but dimmish when they rarely interact.

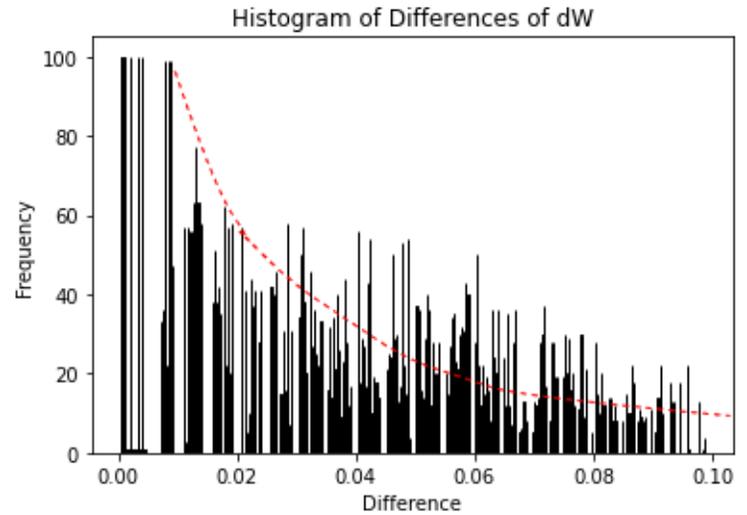

## Second Simulation Setup

Since the first simulation has shown strong interactions with the neurons and their neighbors, I researched this further with the second simulation. The second simulation contains four simulations of 100 neurons each. Only one shows how the weights are randomly distributed. The other three simulations show how communities of neighboring neurons can occur from the randomized initial weights even in a smaller environment of a 10 by 10 grid. Each simulation shows distinct features.

**Smaller Simulations**

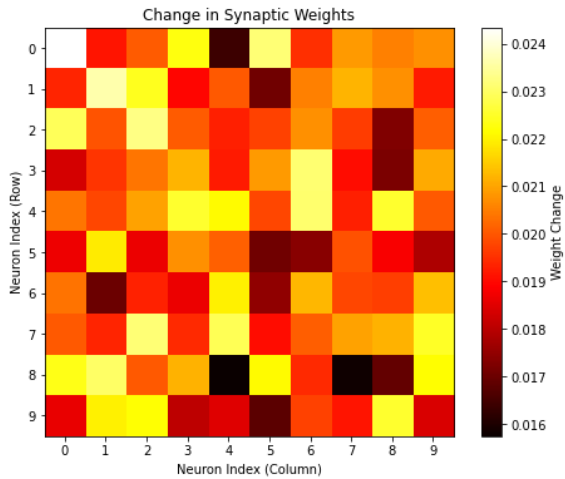

*Figure 10.*

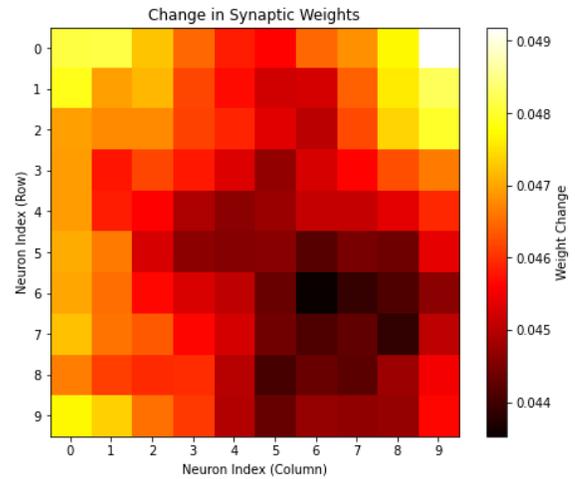

*Figure 11.*

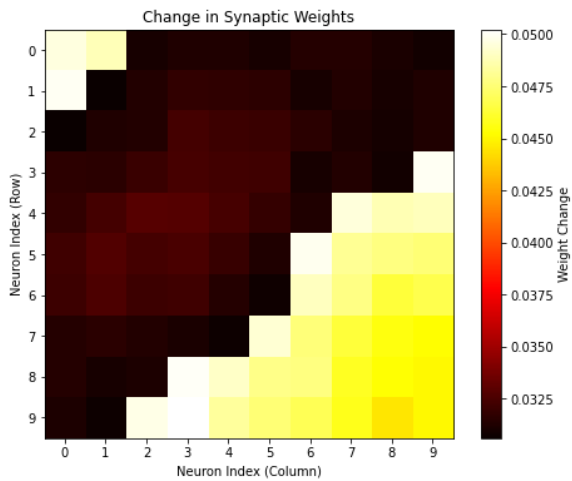

*Figure 12.*

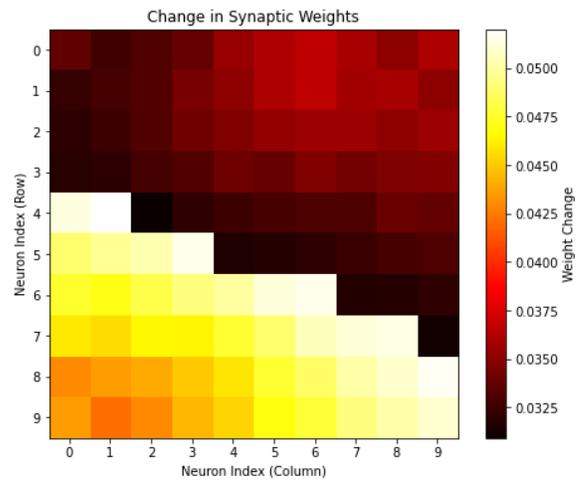

*Figure 13.*

Figure 10 shows the initial randomized weights, where there is no obvious correlation between the neurons and their neighbors. However, the other simulations do show the correlation, where when there are multiple communities that border each other, the weights are extremely high or low on the border. The extreme change in weights shows that it is separate communities (Shown by Figure 12 and Figure 13). Figure 11 shows a single community of neurons and there are no extreme changes in weights.

The figures help support how connected neurons are to their neighbors and separating into different communities at a smaller scale supporting the first simulation's result that the neurons connections to one another are extremely strong. The strength of the connections seen at this scale show how dependent the neurons are to one another, strengthening the claim of biological plausibility of the first simulation.

**Third Simulation Setup**

I simulated each type of SNN with one neuron that I will test over 100ms, this includes Izhikevich model, LIF, and the Hodgkin-Huxley model. Each has its own properties I would like to compare them to GAIN.

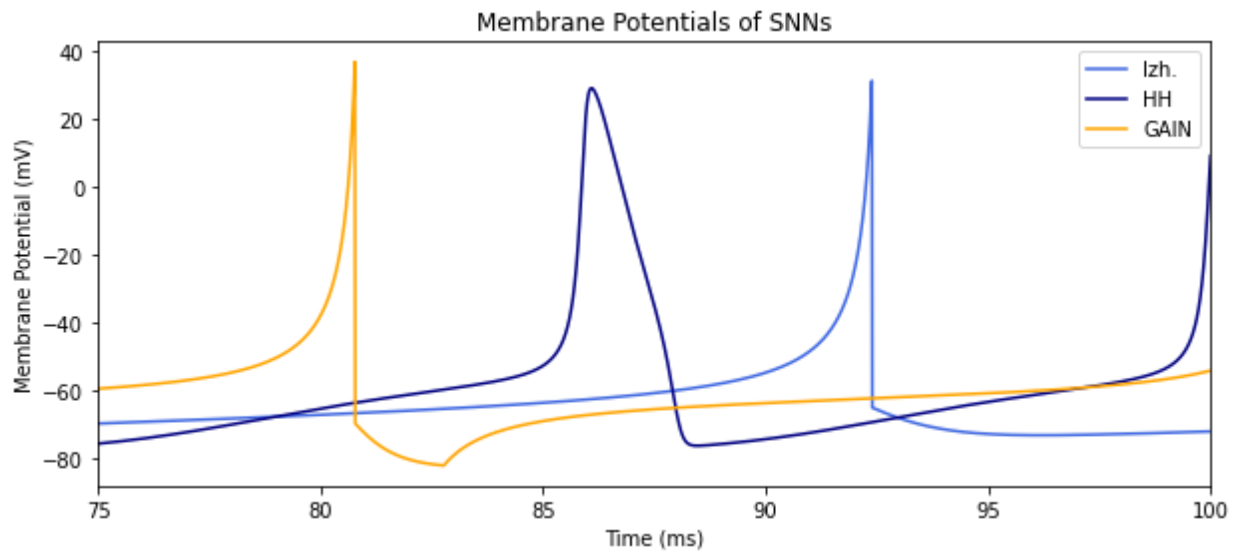

*Figure 9. Shows a 1000ms single neuron simulation of the Hodgkin-Huxley, Izhikevich model, and GAIN. With a timeframe of 75ms to 100ms.*

Figure 9. helps to show how the GAIN model compares to other alternatives to a biological neural network. The Hodgkin-Huxley model in dark blue shows how a biological

neuron would react and shows its behavior. The Izhikevich model in light blue shows how the behavior compares to GAIN with adjustments to its structure. The graph shows that the GAIN model has a major improvement in its behavior including a repolarization period. As it has been mentioned, it was implemented along with a refractory period it compares it to a biological neuron. The graph also shows how the GAIN model kept the Izhikevich model's structure in membrane potential, however, the structure of the model in the entire network is what gives it more biologically realistic activity.

## Conclusion

### Implications

The GAIN model demonstrates promising results in simulating biologically realistic neuronal behavior. It can suggest that the model could serve as a valuable tool for exploring various neural network dynamics, including the impact of network topology, shape, and learning rules on processing information. It can also be useful in different fields such as education, neuroscience, computer science, and artificial intelligence. In addition, because of biological plausibility its applications are heavily focused on pattern recognition. Its uses will continue to expand and evolve as it develops.

### Limitations

The GAIN model still contains issues that will need to be resolved in future development. When I programmed the simulation I programmed it as data-oriented, this means every neuron and its data within arrays. This structure of code makes it difficult to implement features that would be useful to assess the full potential of the model. I was limited by my processing power,

CPU with four cores, and time. As the model increased in complexity so did its processing exponentially, this required me to focus more on the behavior and ensure that its ability to learn more than the processing time, for example, a simulation of 10,000 neurons took 21 hours which is extremely slow. The model also lacks the ability to adjust each neuron separately and has preset variables for distinct types of neurons. Distinct types of neurons have also been left out only focusing on PFC neurons, it would be extremely beneficial to see how it reacts to different systems such as sensory neurons.

s

**The Future**

Future work will focus on integrating more complex stimuli and enhancing the model's capabilities to simulate other types of learning mechanisms. Additionally, further exploring the applications and uses of GAIN, as previously mentioned, can be used for neuroscience and pattern recognition, but as it develops for the future, I could see its applications growing larger. Which could further prove its effectiveness as a computational model of biological neural networks.

Future testing of the GAIN model will be programmed with C++ and make it object-oriented[12] instead of data-oriented, although causing some overhead object oriented will be beneficial for future development treating each neuron separately. A dendrite radius that grows with stimulus can also be added, so neurons can have more than 8 neighbors, this will help further make the model biologically plausible. To complement the dendrite radius, adding a trimming function to get rid of a neuron's weakest connections would be best. This would help with the efficiency of the model and improve its biological plausibility even further. A future goal for the model is to be able to input data and have it normalized for the input current and read the output spikes with intervals of time as binary. This would allow the model to be trained in complex data and give complex outputs, this was one of my original goals for this version of the model, but I was limited by my time constraint.

---

[12] Objected oriented programming, unlike data-oriented programming, organizes code around "objects" that combine data and actions. Each object represents a concept that has its own properties and methods.

**Final Analysis**

The GAIN model, although promising is missing several features that can improve it dramatically as it continues to develop, such as dendrite radius, improving the equations of the model, and the ability to train it. Although it can be trained because of STDP I can see implementation of optimization algorithms such as gradient descent being more effective with it. STDP also makes it extremely difficult to work with, and a few of the future goals include making it easier to work with.

Currently it has proved to be accurate enough to be considered biologically plausible. I also enjoyed the implementation of the grid-based structure. Since it did make it easier to work with and to simulate a MEA, this grid-based structure I have not seen before within any type of neural network before. It will be extremely interesting to see how the structure develops since it can be adjusted to fit into multiple dimensions and does not have the computational overhead of neuron simulation within a space, a type of simulation I was actively trying to avoid. I see this version of the model being an amazing framework to show the potential of what the model is capable of, specializing within a range of fields impossible for traditional neural networks.

# Appendix A

## Structure and Function of the Izhikevich Model

The structure of the Izhikevich Model is a membrane potential function that is dependent on the recovery variable. While the recovery variable is dependent on two other variables both "a" and "b" which is what largely contributes to the dynamics of the model. Where "a" is the decay rate and "b" is the sensitivity of model. The functions of the membrane potential and recovery variable are shown by (1).

$$V' = 0.04V^2 + 5V + 140 - u' + I \tag{1}$$

$$u' = a(bV - u) \tag{1.1}$$

The Izhikevich model also requires a function to reset the membrane potential and recovery variable if it hits a maximum voltage typically used as 30mV as shown by (2). To prevent the exponential growth of the membrane potential. "I" is the input current which is typically a constant value.

$$\text{If } V \geq 30\text{mV, then } \{V \leftarrow c \quad u \leftarrow u + d \tag{2}$$

Equation (2) uses "c" and "d" as the reset variables where "c" is the value which the membrane potential resets to, and "d" is the reset variable of the recovery potential. Allowing for a simplistic and relatively dynamic model that can replicate the biological mechanisms seen in biological neurons such as bursting and spiking.

## Appendix B

## Structure of the GAIN model

GAIN is an altered Izhikevich model where its structure is a grid where each neuron connects to its closest neighbors (See Figure 1). This structure is to make sure that the model encourages the adaptation of the network to any stimulus and encourages synaptic plasticity. To improve the dynamics of the Izhikevich model, the decay variable "a" as a function of the input current and the sensitivity variable "b" as a function as shown by (2).

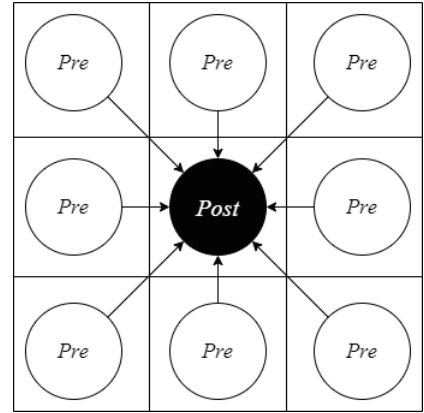

*Figure 1*

$$a(I) = a_0 + \sigma I \qquad (2)$$

| | | | |
|---|---|---|---|
| | | $b(V) = b_0 + \gamma V$ | (2.1) |
| | | | |

Where $a_0$ is the initial decay variable and $\sigma$ is a constant that defines how dependent the function a(I) is on the input current. While $b_0$ is the initial sensitivity variable and $\gamma$ is a constant that defines how dependent the function b(V) is on the membrane potential. The main functions are still similar besides with a few exceptions particularly with the recovery variable as shown by (3)

$$V' = 0.04V^2 + 5V + 140 - u' + I' \qquad (3)$$

| | | | |
|---|---|---|---|
| | | $u' = a(I)(b(V) - u)$ | (3.1) |
| | | | |

The input current is a function of time, and the sum of a neuron's neighbors' synaptic weights by the change of the neighbors' membrane potential as shown by (4).

$$I' = \left( \sum W_{syn} \times \Delta V \right) + I \qquad (4)$$

| | | | |
|---|---|---|---|

The synaptic weight of a neuron is determined by the initial weight and the change of the weight, multiplied by STP (See Appendix D). I adjusted the effective weight formula to still account for STDP within a smaller time, as shown by (5), and the effective weight is the synaptic weight for each neuron.

$$W_{syn} = (W_0 + \Delta W) \times u(t) \times R(t) \tag{5}$$

$W_0$ is the initial weight for a neuron, $\Delta W$ is the STDP weight (See Appendix C). The synaptic weight describes the effect a neuron has on its neighbors. The synaptic weight is what improves the interactions within the entire network.

# Appendix C

## Spike-Timing Dependent Plasticity

STDP is needed to adjust the weights of connections to a neuron and improve learning. With repeated inputs to a neuron, it can decrease or increase the resistance that other neurons affect it as time progresses and increases ($\Delta t$).

$$\Delta W = \{A_+ \times e^{\frac{-\Delta t}{\tau_+}}, \quad \Delta t > 0 \quad -A_- \times e^{\frac{\Delta t}{\tau_-}}, \quad \Delta t < 0 \tag{1}$$

The hard bounds, $A_+$ and $A_-$, are an update rule to keep the weights within a range. With them being defined by a sigmoid function (S), the maximum weight ($w_{max}$), and soft bounds ($\eta_+$ and $\eta_-$) as shown by (2)

$$A_+ = S(w_{max} - w)\,\eta_+ \tag{2}$$

$$A_- = S(-w)\,\eta_- \tag{2.1}$$

Sigmoid function, as shown by (3), allows for a smooth transition for any weight in the range of the soft bounds. A sigmoid function helps decrease fluctuation and keep weight changes as minimal and decreases the chance of overshoot. Where k is a constant which increases or decreases the rate of change.

$$S(x) = \frac{1}{1 + e^{-kx}} \tag{3}$$

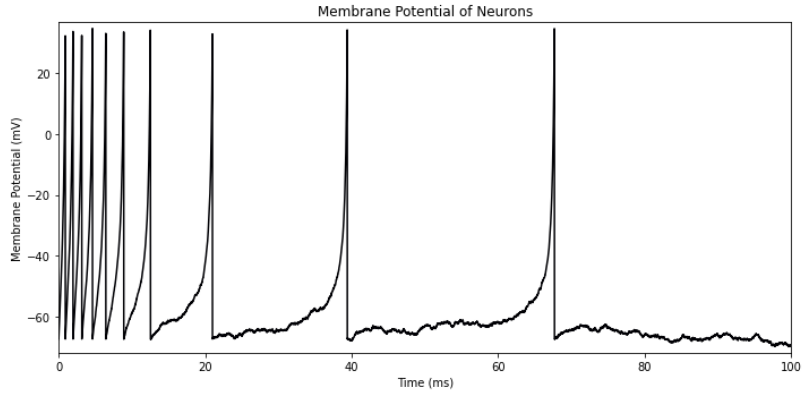

*Figure a. A single neuron simulation, with a constant 10mV stimulus*

Figure a. shows a primitive version of the model missing several features, including a refractory period, but demonstrating STDP. The number of spikes decreases with time, it shows how the neuron becomes less reactive to the stimulus. The decreased number of spikes over time demonstrates the potential for the model to learn.

# Appendix D

## Short-Term Plasticity and Weights

STP is the change in synaptic strength based on the timing of previous spikes. Which can be represented by facilitation and depression, which together make up the effective STP as shown by (1) and (2).

$$u(t) \ = \ u_0 + U(1 - u_0) \times e^{\frac{-\Delta t}{\tau_f}} \tag{1}$$

$$R(t) = R_0 \times e^{\frac{-\Delta t}{\tau_d}} + 1 - e^{\frac{-\Delta t}{\tau_d}} \tag{2}$$

While U is a constant that has a small variable that represents a small membrane potential. The smallest membrane potential that U can be is zero, and this can be seen as the sensitivity of the variable.

The effective weight of a single neuron is the initial weight, sum of the weights, and the effect that STP has on it. The initial weights can be adjusted, through backpropagation or another optimization algorithm, to train the model. The effective weight being represented by (3).

$$W_{eff} = \ W_0 \times u(t) \ \times R(t) \ + \ \Delta W(t) \tag{3}$$

Equation (3) shows that the effective weight is STP by LTP/LTD and added by the sum of the weights. This is simple, but important for implementing STP into the model. The uses of STP are for increasing the dynamics and biological plausibility of the model efficiently. This allows neurons to affect and interact with their neighbors (As Shown by Figure a.)

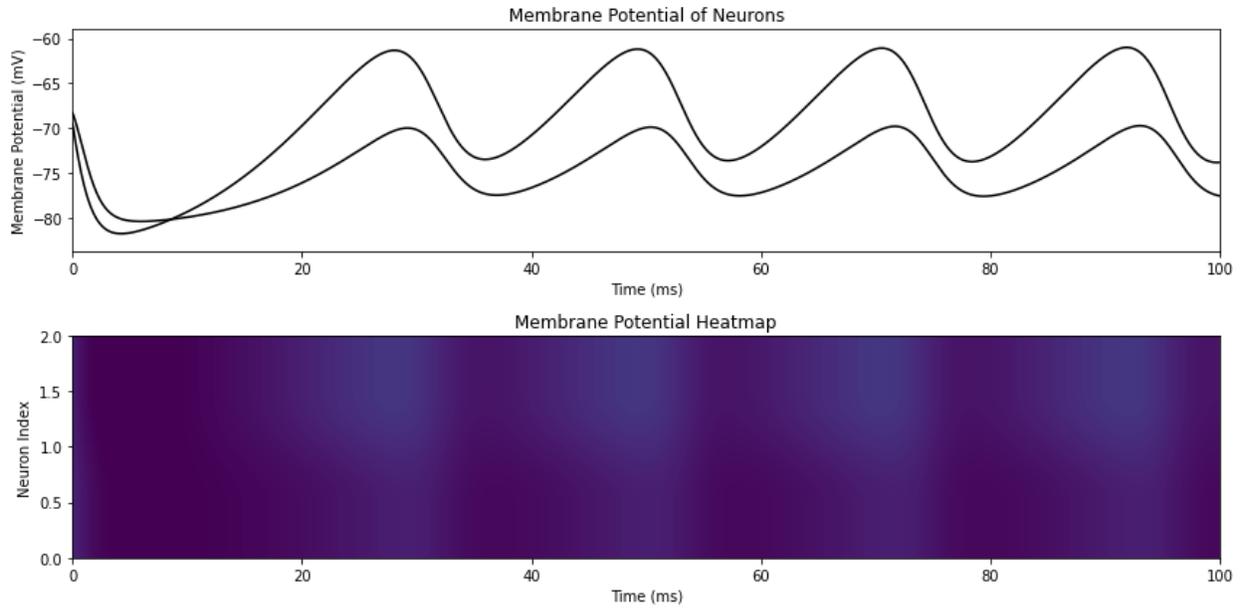

*Figure a. A stimulus based on a sines function applied to neuron two only. The behavior from neuron one is from the interactions with neuron two.*

# Appendix E

## Table of Variable Ranges

| *Variable* | Lower Bound | Upper Bound |
|:---:|:---:|:---:|
| $a_0$ | 0.02 | 0.1 |
| $b_0$ | 0.2 | 0.5 |
| $\sigma$ | 0 | 0.1 |
| $\gamma$ | 0 | 0.1 |
| $\tau_\pm$ | 10ms | 40ms |
| $\tau_-$ | 10ms | 40ms |
| $\tau_f$ | 10ms | 100ms |
| $\tau_d$ | 50ms | 300ms |
| $U$ | 0 | 0.1 |
| $R_0$ | 0 | 0.2 |
| $u_0$ | 0 | 0.2 |
| $W_0$ | -0.01 | 0.5 |
| $\eta_\pm$ | .001 | .01 |
| $\eta_-$ | .001 | .01 |
| $k$ | 0 | 1 |